\documentstyle[11pt,a4wide,epsf]{article}
\newcommand{\x}{\raisebox{0.08cm}{\mbox{$\chi\ $}}}
\newcommand{\wn}{\mbox{$W_\nu$}}

\newcommand{\ec}{Euler-characteristic}

\newcommand{\mf}{Minkowski-Functionals}

\def\R{\mbox{\rm I\kern-.18em R}}
\def\E{\mbox{\rm I\kern-.18em E}}

\catcode`\@=11
\long\def\@makecaption#1#2{%
   \vskip 10\p@
   \setbox\@tempboxa\hbox{{\bf#1:} #2}%
   \ifdim \wd\@tempboxa >\hsize
       {{\bf#1:} #2}\par
     \else
       \hbox to\hsize{\box\@tempboxa\hfil}%
   \fi}
\catcode`\@=12
\begin{document}

\title{\bf Applications of Minkowski-Functionals to 
    the Statistical Analysis of Dark Matter Models}

\author{{\bf Michael Platz\"oder}\\  Max--Planck--Institut f\"ur Astrophysik\\ 
Postfach 1523, D-85740 Garching, F.R.G.\\{}\\
{\bf Thomas Buchert}\\  Theoretische Physik, 
Ludwig--Maximilians--Universit\"at\\Theresienstr. 37, D-80333 M\"unchen, F.R.G.}  

\maketitle
%%%%%% abstract  %%%%%%%%%
\begin{abstract}
\noindent
A new method for the statistical analysis of 3-D point processes,
based on the family of \mf\ , is explained and applied to modelled
galaxy distributions
generated by a toy-model and cosmological simulations of the 
large-scale structure in the Universe. These measures are
sensitive to both, geometrical and topological properties of spatial
patterns and appear to be very effective in discriminating
different point processes. Moreover by the means of conditional
subsampling, different building blocks of large-scale structures like sheets,
filaments and clusters
can be detected and extracted from a given
distribution. 

\end{abstract}

%%%%%% Introduction  %%%%%%%%%

\section{Introduction }
\label{1_introduction}

The still unsolved question of the origin and evolution of the 
Large-Scale Structure (LSS) of the universe is a central topic 
in modern cosmology. The scientific approach to this problem is based on three
columns:
\begin{enumerate}
  \item The distribution of matter in the Universe is assumed to be
	traced by luminous galaxies. Thus, observation of galaxies
	and measurement of their redshifts is, besides peculiar--velocity 
        measurements, the only way 
	to gain an idea of the matter distribution in our local
	area of the Universe.
	
  \item Todays powerful computer systems make the numerical simulation of
	statistical ensembles of theoretical models 
        of structure formation possible. By this
	way effects of different dark matter models on the resulting structures
	can be studied.

  \item The comparison of observations with theoretical models is made
	in terms of statistical measures which can be applied to
	both, real and simulated galaxy distributions. Thus, models can
	either be sorted out or favoured and improved to fit  
	observational results.
\end{enumerate}

The introduced statistical method belongs to the third column and
offers a new approach for finding adequate measures which are 
capable of discribing
and characterizing global and local features of galaxy distributions.
It was first suggested by Mecke, Buchert and Wagner \cite{mec94},
but only now numerical problems are overcome and first results can be gained.

Popular measures in that field include {\em N-Point Correlation Functions}
\cite{pee80}, {\em Counts-in-Cells} \cite{gaz93}, {\em Void
Probability Functions} \cite{whi79}, {\em Percolation Analysis}
\cite{zel82}, {\em Minimal Spanning Trees} \cite{bar85},
{\em Genus of Isodensity Levels} \cite{got86}, {\em Voronoi-foam
Statistics} \cite{wey87, wey91} and many more.
 
After discussing the basic properties of \mf\ in Section
\ref{2_minkowski}, the new statistical method is presented in Section \ref{3_method}. 
Its sensitivity to different components of the LSS (clusters, walls,
filaments) is investigated, using toy-models based on
Voronoi-tesselations, and the selection of structures by
the means of {\em conditional subsampling} is illustrated (Section \ref{4_toy}).
Finally the method is applied to a series of CDM-simulations in
Section \ref{5_sim}, a short summary is given in Section 
\ref{6_con}, and an outlook on future prospects in Section \ref{7_out}.

%%%%%% Section 2 %%%%%%%%%%%%%%%%%%%%%%%%%%%%%%%%%%%%%%%%%%%%%%%%%%%%%%%%%%%%%%%

\section{Minkowski-Functionals}
\label{2_minkowski}

The \mf\ \wn\ have their origin in the mathematical theory of convex bodies
and integral geometry\footnote{A more detailed discussion of
  the \mf\ can be found in \cite{mec94}, \cite{mec93} and \cite{pla95}. The
  mathematical background of integral geometry is covered by
  \cite{had57}.}.
In $d$ dimensions there exist $d+1$ of these
functionals including {\em geometrical} and {\em topological}
descriptors to characterize content (volume and surface), 
shape and connectivity
of a body $A\subset \R^d$ :

In the case of three dimensions we have: 
  \renewcommand{\arraystretch}{2}\boldmath
  \begin{center} \setlength{\fboxsep}{10pt} 
  \fbox{ $\begin{array}{rcll}
		W_0 & = & V(A)      & \quad\mbox{(volume)}\\
	       3W_1 & = & S(A)      & \quad\mbox{(surface)} \\
	       3W_2 & = & H(A)      & \quad\mbox{(shape)}\\
	       3W_3 & = & G(A) \\
                    & = & 4\pi\x(A) & \quad\mbox{(connectivity)}\\
		\end{array} $ }\unboldmath 
   \renewcommand{\arraystretch}{1}
   \end{center}\setlength{\fboxsep}{4pt} 
\unboldmath 

Here, shape is expressed in terms of the integral mean curvature
$H$ of a body's surface, and the integral Gaussian curvature $G$, 
related to the Euler-characteristic $\chi$ via the Gau{\ss}-Bonnet
theorem, is a measure for
the connectivity.

Important properties of the \mf\ include: 

\begin{tabular}{lrcl}
 \mbox{\bf Motion Invariance :}& $\wn(uA)$&=&$\wn(A)$\\
 \mbox{\bf Additivity :}   & $ \wn(A\cup B)$&=& $\wn(A)+\wn(B)-\wn(A\cap B)$
\end{tabular}

Thus, the measures are invariant against translations and rotations of
the body or combinations $u$ of the two. The \wn\ of a body $C$, which is the
union of two point sets $A$ and $B$, obey a simple additivity relation
that can be extended by induction to an arbitrary number of components
of $C$. Except for the volume, all other measures are located on the
surface of the considered body.

\subsection*{\ec}

The \ec\ $\x=G/4\pi$ of a body $A$ is related to its
 {\em genus} $g$ 
\[ \x(A) = 1 - g(A) \;\;,\] 
so in three dimensions \x can be expressed by :
\[ \x = components - tunnels + cavities \;\;. \]
In this way it can be used as a {\em topological} measure for the connectivity 
of a point set.

\begin{figure}[ht]
  \centering 
  \fbox{\epsfxsize=14.0cm \epsffile{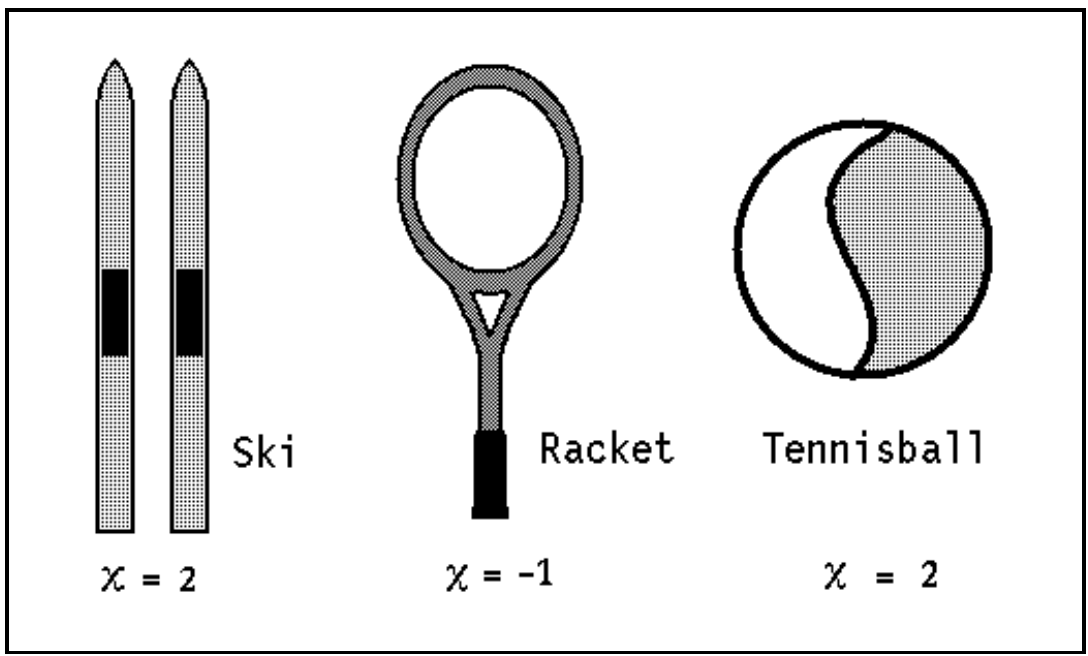}}
	\renewcommand{\baselinestretch}{0.75}
	\caption{\label{PIC_sports} \footnotesize 
         {\bf Example:} A pair of ski consists of two components and, 
         unless there are
         holes or cavities inside a ski, an {\ec} of $\chi=2$ is found. 
         The tennis
         racket has two "tunnels" and one component yielding $\chi=-1$
         (however, a more realistic one would have a highly negative $\chi$). 
         Finally a tennis ball has a cavity and thus $\chi=2$.}
	 \renewcommand{\baselinestretch}{1.0}
 \end{figure}
	
%%%%%% Section 3 %%%%%%%%%

\section{Method}
\label{3_method}

\subsection{Boolean Grain Model}

\begin{figure}[ht]
  \hspace{1cm}  \epsfxsize=12cm \epsffile{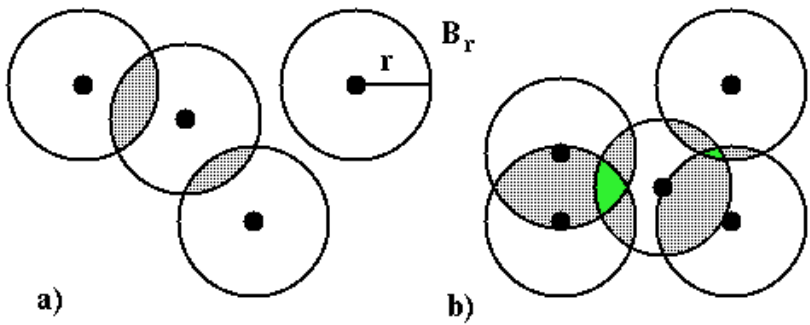}

%\vspace{-0.5cm}

  \caption{\label{PIC_balls} \footnotesize Each point is decorated with
                 a ball $B_r$. }
\end{figure}

In a given distribution the points (galaxies) are represented by their
cartesian coordinates $\vec{x}_i$.
The input data are the
coordinates of all the $N$ points inside a cubic volume $\cal C$ that is cut out of
the point distribution. This cube is then scaled to an edgelength of one
(unit cube ${\cal C}_u$). Outside this box periodic boundary conditions are
used, i.e. ${\cal C}_u$ is supplemented by 26 identical boxes around it.
For the calculations, on every point a ball $B_r$ with radius $r$ is
centered (Boolean grain model).

The Minkowski measures of the point set 
$ {\cal B}(r)=\cup_{i=1}^N B_r(\vec{x}_i) $
which consists of all the points inside the union of the balls are
then calculated, partly using numerical methods. Actually the {\em local} 
contributions $\wn(r,\vec{x}_i)$ to the functionals from the surface of each single ball
$B_r(\vec{x}_i)$ (i=1,...,N) are determined and {\em global} measures 
arise by taking the mean of all the $\wn(r,\vec{x}_i)$ for a
given radius $r$. It can be shown that these global results contain 
$n$-point correlations of every order $n$.

The radius $r$ of the balls is the parameter that is needed to analyze
the distribution on different scales. For small radii most of
the balls are isolated, thus yielding the measures of a ball ($\x=1$). With
growing $r$ more and more balls intersect and connect to network-like structures
($\x<0$). Finally, the network turns into a body with enclosed
cavities ($\x>0$) which then are filled when the cube ${\cal C}_u$ ist
totally overlapped by ${\cal B}(r)$.

%\vspace{-1.5cm

\begin{figure}[ht]
    \hspace{1cm}\epsfxsize=13.0cm \epsffile{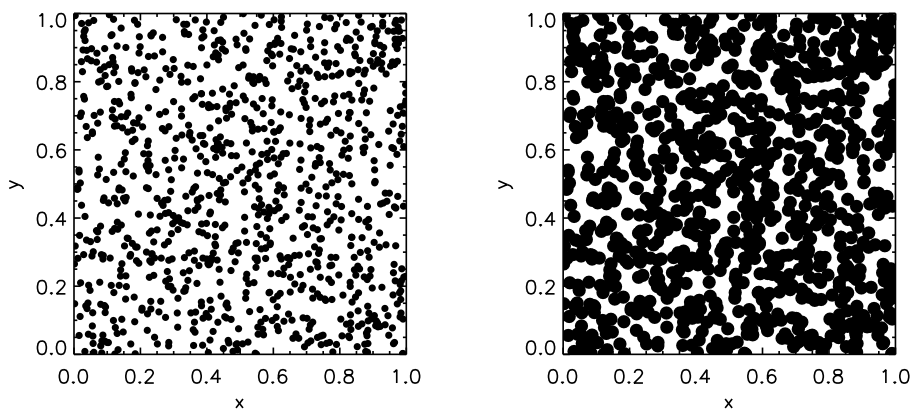}

%\vspace{-0.5cm}

    \caption{\label{IDL_modell}\footnotesize
     Boolean grain model for a Poisson process for two values of $r$,
     corresponding to ($x=r/{<D>}$) 
     $x_1 = 0.18$ (left) and $x_2 = 0.36$ (right);
     $<D>$ is the mean distance of points in the set.}
\end{figure}

\subsection{Results for Poissonian Sample }

The mean values of the \mf\ can be calculated analytically for a
stationary Poisson process \cite{mec91}. The radius is scaled by the
mean distance $<d>$ between the points in the cube ${\cal C}_u$.

\vspace{-2cm}

\begin{figure}[ht]
  \hspace{1cm}  \epsfxsize=11cm \epsffile{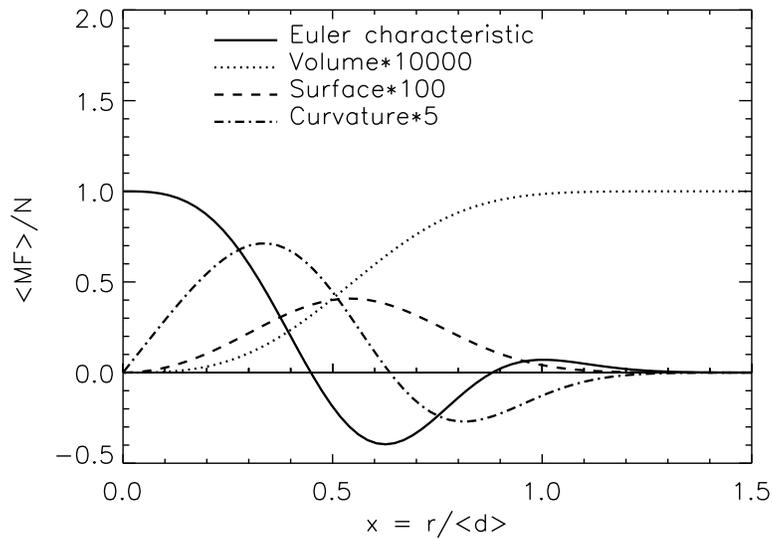}

\vspace{-0.5cm}

  \caption{\label{IDL_pois}  \footnotesize {\em Global} MF per ball 
                            (analytical).}
\end{figure}

\begin{figure}[ht]
    \hspace{1cm}\epsfxsize=12.0cm \epsffile{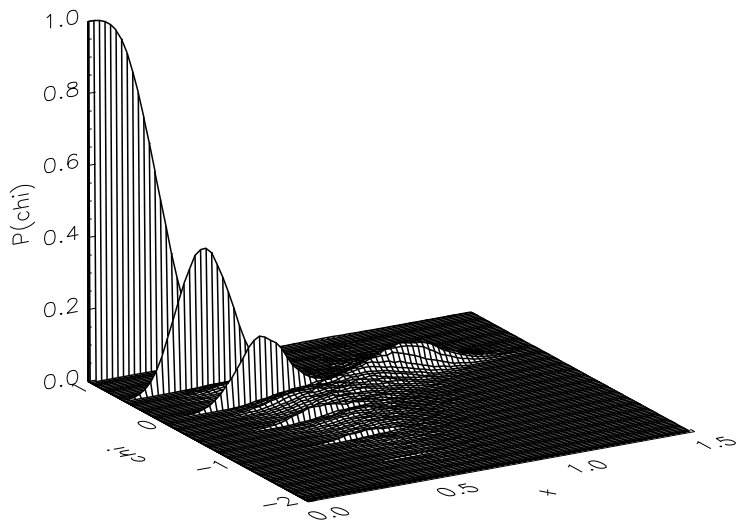}
 
\vspace{-0.5cm}

  \caption{\label{IDL_poisvert} \footnotesize The distribution of the
    {\em local} 
    Euler-characteristic.}
\end{figure}

%%%%%% Section 4 %%%%%%%%%

\section{Analysis of Toy Models }
\label{4_toy}
\subsection{Voronoi Tesselations}

A toy model for the artificial generation of structures, roughly
similar in to the ones observed in the Universe, is based on 
{\bf Voronoi Tesselations}:
\medskip
\begin{figure}[ht]
  \centerline{\epsfxsize=12.5cm \epsffile{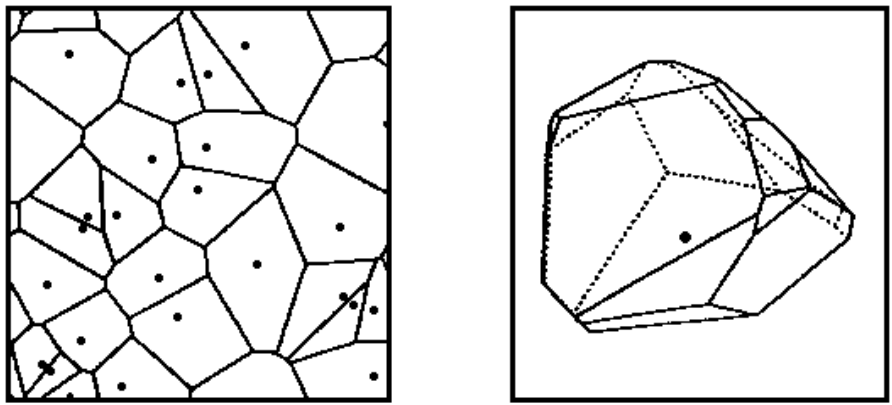}}
  \renewcommand{\baselinestretch}{0.75}

  \caption[]{ \footnotesize (from \cite{wey91}) 
             A {\bf Voronoi-Tesselation} is a decomposition of space
             into convex, non-overlapping
	     domains, which are generated as {\em Wigner-Seitz} cells
	     of a discrete distribution of points;\\
	     {\bf a)} Voronoi-Tesselation in two dimensions,
	     {\bf b)} example for a cell and its nucleus (d=3).}
  \renewcommand{\baselinestretch}{1.0}
\end{figure}
\medskip

A Poissonian distribution of points ("nuclei") inside ${\cal C}_u$
defines a pattern of {\em Wigner-Seitz-cells}, the tesselation.
In a second step a number of $N$ test-points is randomly distributed
inside ${\cal C}_u$. Then these points are projected onto the walls, or edges,
or vertices of the tesselation, thus generating sheets (or walls), filaments 
or clusters. For all these structure components one can choose a
finite thickness and the ratio of points ending up in each of them. 

\subsection{Walls, Filaments and Clusters}

The method is sensitive to different structures in a distribution of
points. This can be illustrated by looking at the local contributions 
to the \ec\ \x on the surface $S$ of each ball. As \x is equivalent to the
integrated Gaussian curvature of the considered body,
it can be split into three contributions\cite{mec94}, coming from the uncovered surface
($\chi_{Sur}$), the intersection lines ($\chi_{Lin}$) of two balls
on $S$ and the triplepoints ($\chi_{Tri}$) where three balls have
a common point on $S$. Surface points, that belong to more than three
balls have a low probability and can be neglected.

While for walls, \x includes contributions from lines and
triplepoints over a wide range of $x$, hardly any triplepoints appear for
filaments. The \ec\ \x has a narrow minimum for clustered distributions,
but no distinct minimum for filaments.

\newpage
\begin{figure}[ht]
  \hspace{1cm}\epsfxsize=14cm \epsffile{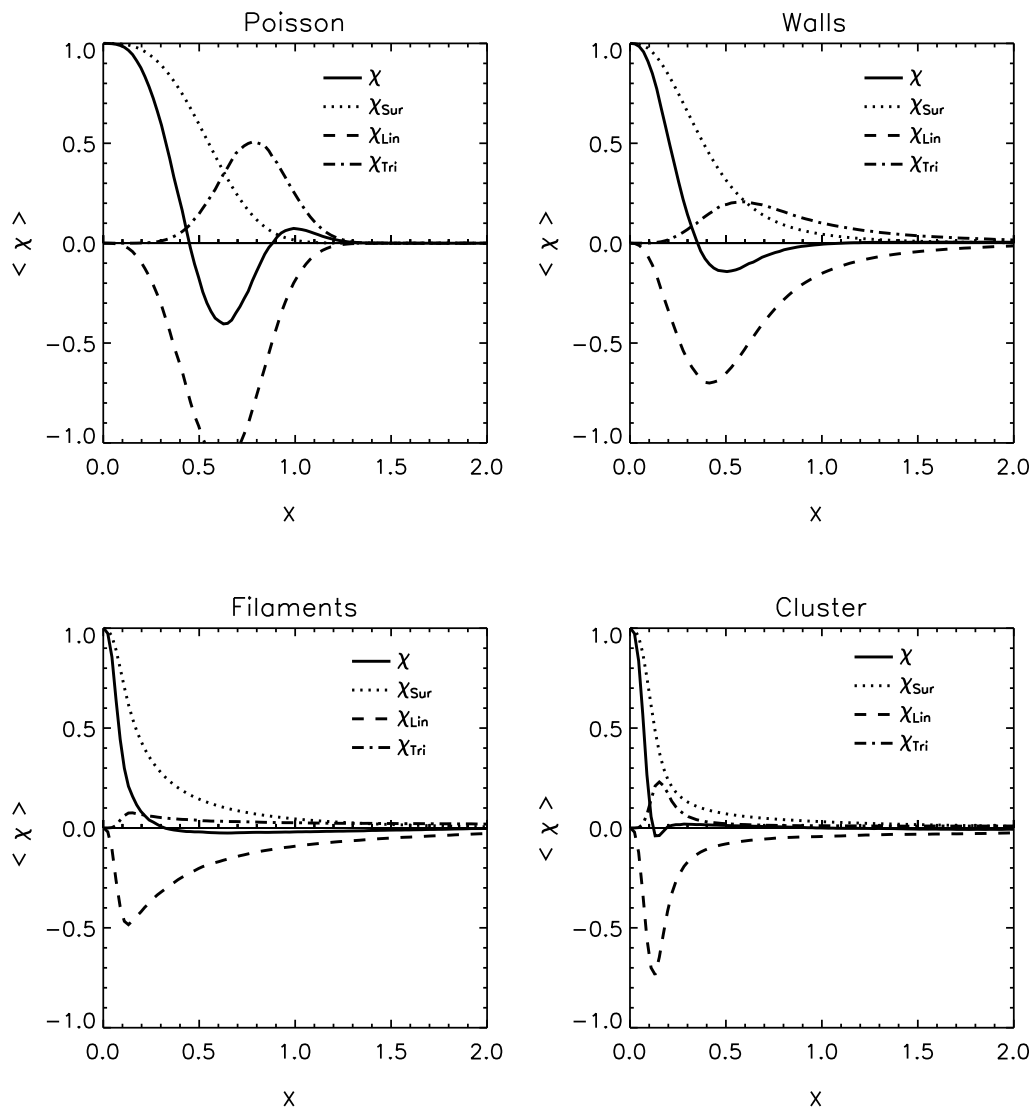}
   \renewcommand{\baselinestretch}{0.75}

   \caption[]{\footnotesize \label{IDL_localrwfc} Local contributions to $\chi$ 
	from the surface ($\chi_{Sur}$), the intersection lines ($\chi_{Lin}$)
	and the triplepoints ($\chi_{Tri}$); ($x=r/{<D>}$).}
   \renewcommand{\baselinestretch}{1.0}
\end{figure} 

%\newpage
\subsection{Conditional Subsampling}

Structures can be detected and analyzed by selecting only 
 galaxies with a local contribution to a functional inside a given
interval.

As an example we consider a mixture of filaments ($4000$ points) 
generated in the Voronoi model and a Poissonian distribution
(also $4000$ points).
We can extract the filament component by suitable conditions on 
the local contributions to the Minkowski-Functionals:
Typical for filaments, e.g., is $\chi\approx 0$ 
(see Figure \ref{IDL_localrwfc}and Figure 8); 
we have chosen the condition
$-0.1\le\x\le 0.1$.

\newpage
\begin{figure}[ht]
  \epsfxsize=16cm \epsffile{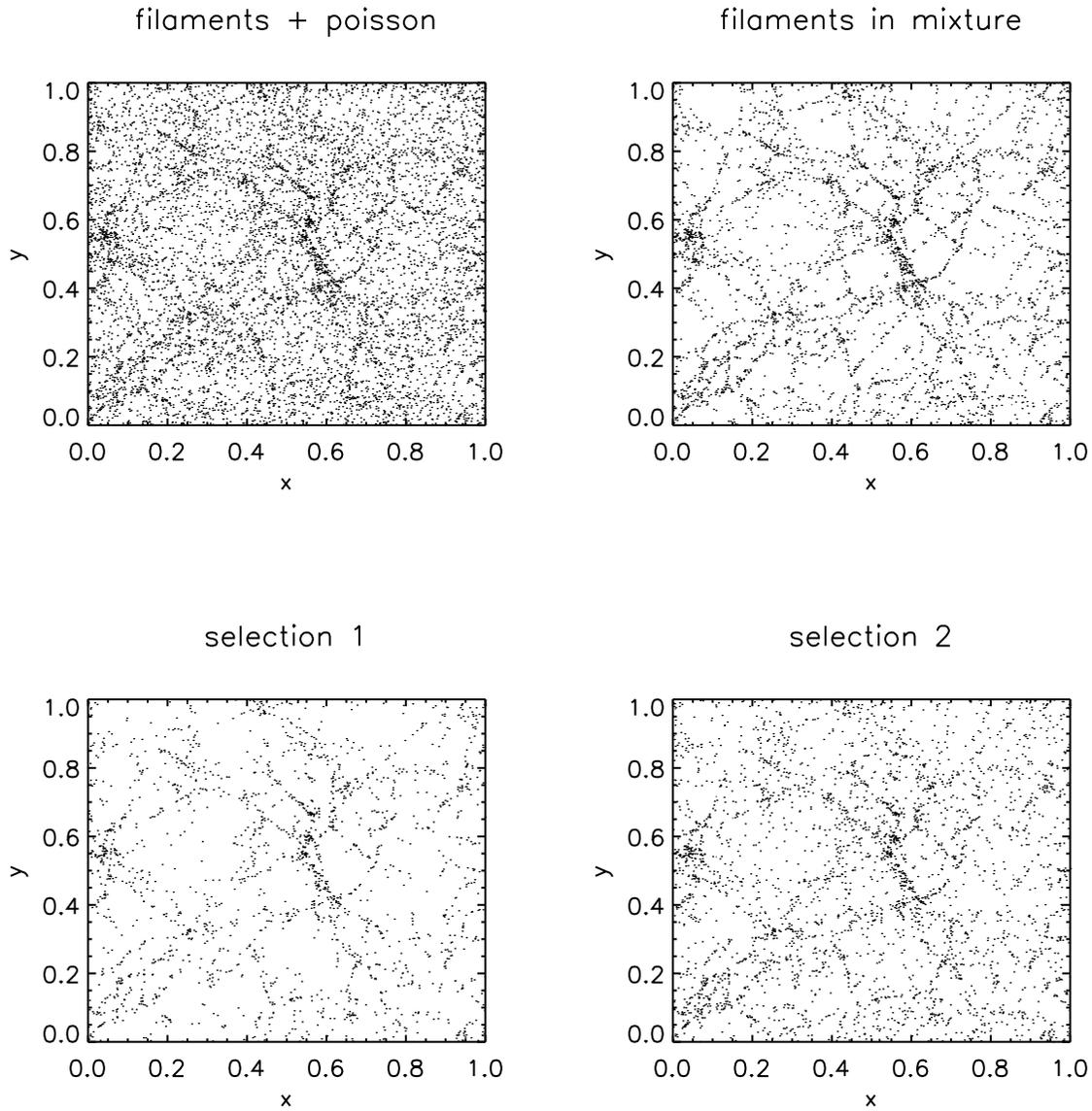}
   \renewcommand{\baselinestretch}{0.75}
   \caption[]{\footnotesize Original distribution, filament component
              and two selected samples at different radii $r$ (corresponding to
              ($x=r/{<D>}$) $x_1 = 0.28\;; x_2 = 0.48$. The
	      first selection consists of about $2000$ points with more
	      than 88\% of them belonging to filaments. The second has
	      3400 points with 70\% in filaments. The pictures show
	      all points of the samples projected onto the x--y plane of
	      the cube ${\cal C}_u$. }
   \renewcommand{\baselinestretch}{1.0}
\end{figure}	

\newpage

\subsection{Two-point distributions with similar two-point correlations}

\vspace{-1.5cm}

\begin{figure}[ht]
  \parbox[b]{8.6cm}{\epsfxsize=8.5cm \epsffile{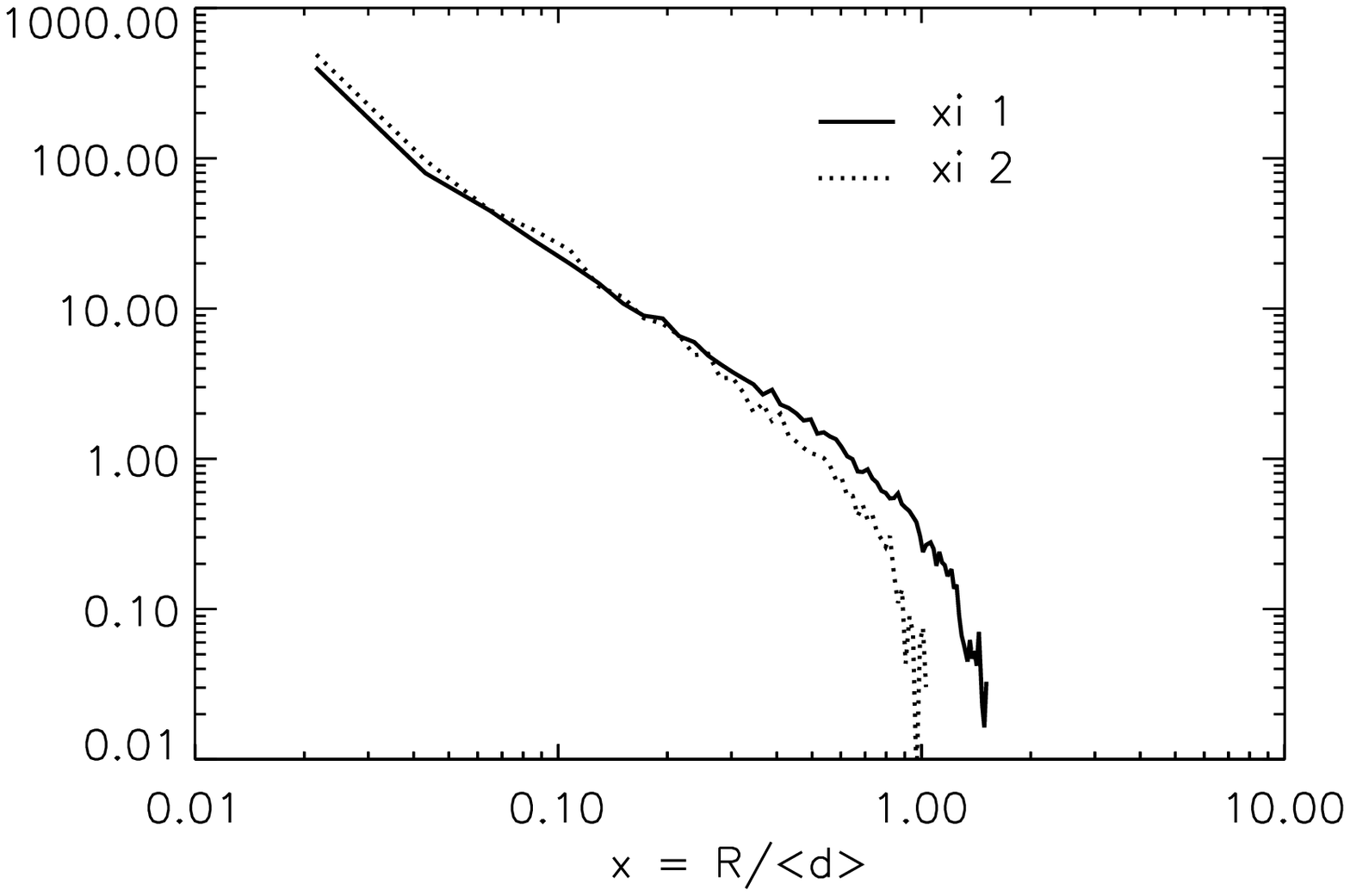}}\hfill
  \parbox[b]{8.3cm}{\epsfxsize=8.2cm \epsffile{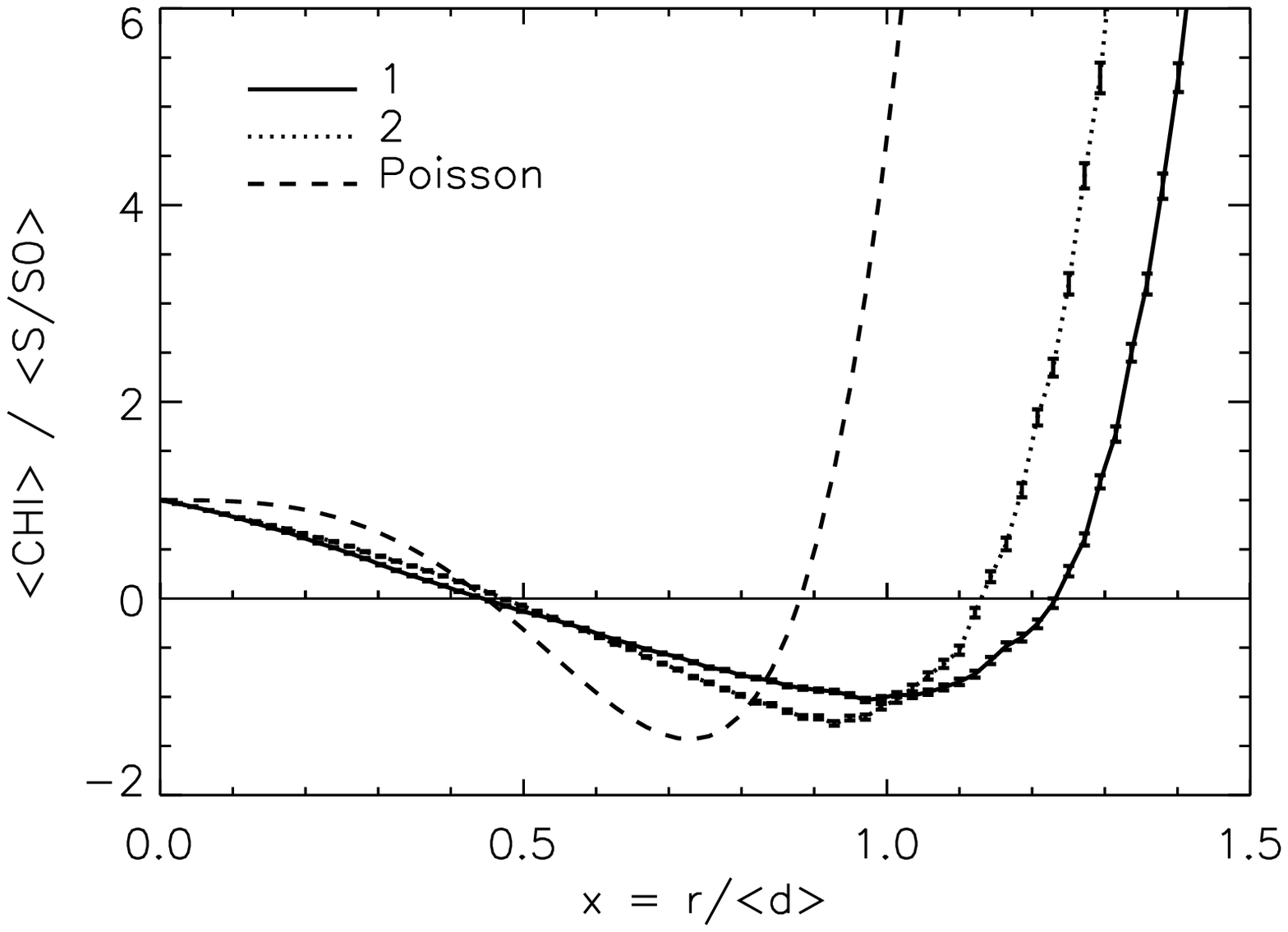}}
   \renewcommand{\baselinestretch}{0.75}
\vspace{-0.5cm}
  \caption[]{\footnotesize The quotient of \x and the uncovered ratio
             of the surface $S$ is a good measure to discriminate the
	     two distributions $1$ and $2$ having similar 2-point
             correlation functions $\xi(x)$ on scales smaller than
	     $x_0$: $\xi(x_0)=1$. The error
	     bars shown are the standard deviations of the mean of the ball
	     contributions in each sample.}
   \renewcommand{\baselinestretch}{1.0}
 \end{figure}	
	
%%%%%% Section 5 %%%%%%%%%

\section{Analysis of CDM-Simulations }
\label{5_sim}

\begin{figure}[ht]
  \hspace{2cm}\parbox{14.0cm}{\epsfxsize=11.0cm \epsffile{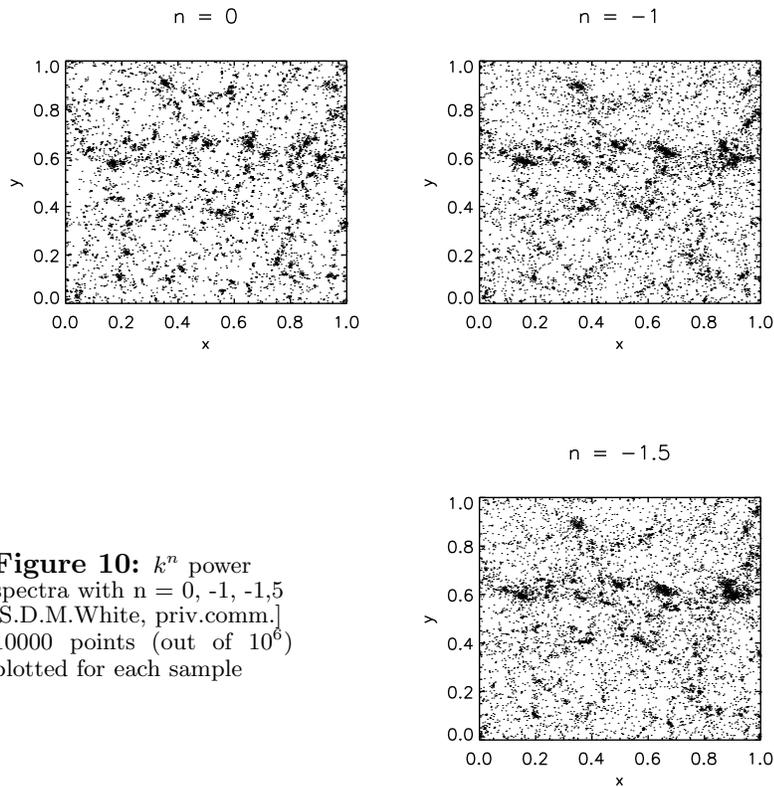}}
  \raisebox{-4cm}{  
\hspace{-11cm} \parbox[b]{4cm}{\renewcommand{\baselinestretch}{0.75}
                  \caption[]{\footnotesize \raggedright 
                  $k^n$ power spectra  with n = 0, -1, -1,5
                  [S.D.M.White, priv.comm.] 

                10000 points (out of $10^6$) plotted for each sample
		   \renewcommand{\baselinestretch}{1.0}}}}
   \end{figure}
\newpage	
The results show clearly the different reaction of the
Minkowski-Functionals to the three spatial patterns.
For a constant power spectrum $\propto k^0$ many small clusters can form,
while for negative $n$ bigger structures appear. In the case
of $n=0$ the ratio of galaxies in clusters is higher than for $n<0$,
so the volume and surface measures yield lower values, while
the mean curvature remains positive. For negative $n$ the points
outside of clusters are part of a network generated by the
corresponding balls in the grain model, thus adding high surface contributions
and negative curvature values (curvature of intersection lines
$H_{Lin}<0$ !).

\begin{figure}[ht]
  \epsfxsize=16cm \epsffile{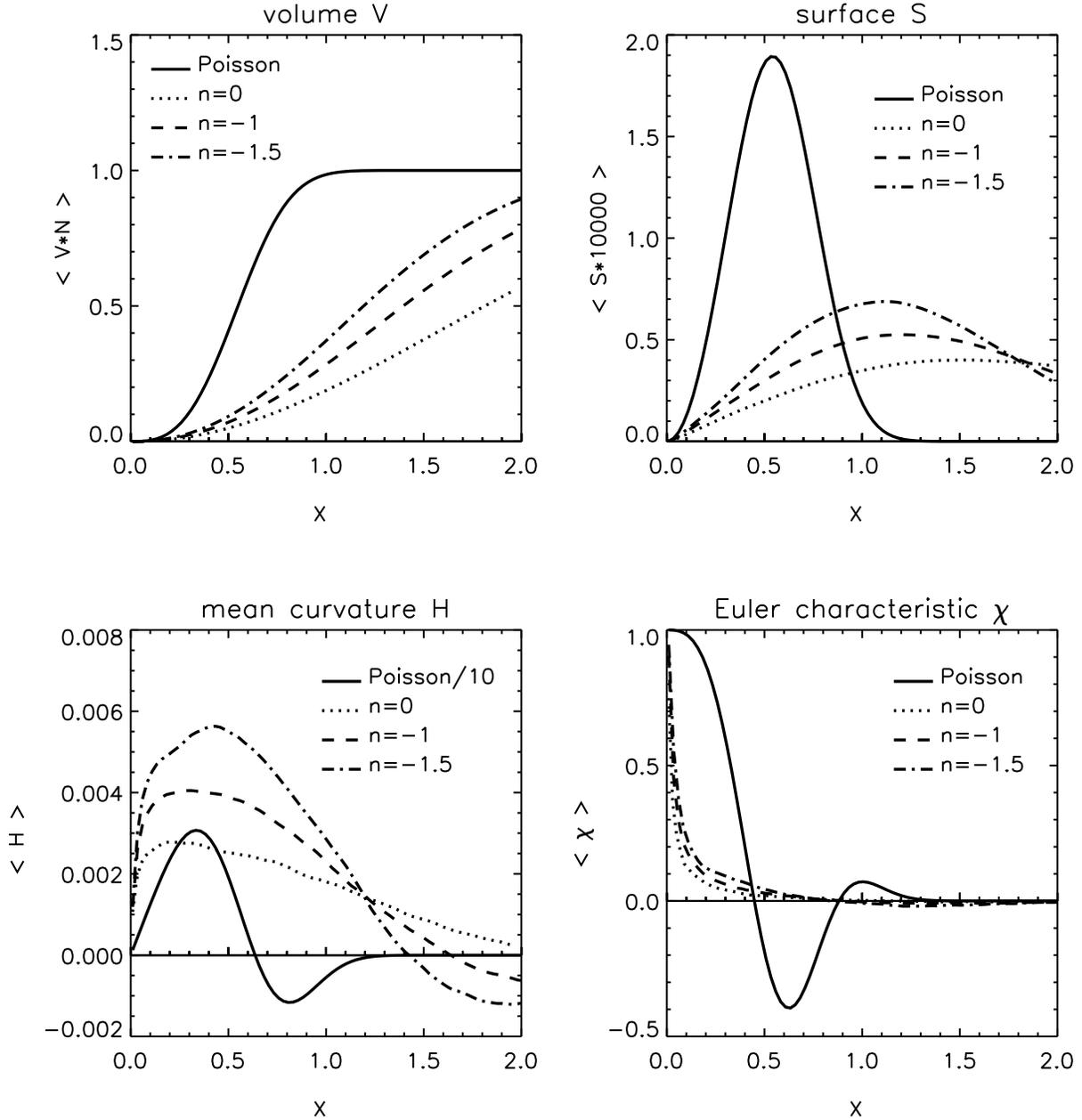}
	\caption{\footnotesize Minkowski-Functionals for CDM
simulations with $k^n$ power spectra; ($x=r/{<D>}$).  }
 \end{figure}	

%%%%%% Section 6 %%%%%%%%%

\section{Conclusions }
\label{6_con}

The introduced statistical method is based on a solid mathematical
background and no special assumptions about the sample have to be made.
The mean values of the functionals over the balls in a sample
are statistically unbiased descriptors and include topological
{\em and} geometrical measures, containing n-point correlations 
of every order\cite{mec94}.
They are statistically robust even for small samples and depend on the radius of
balls which is the (single) diagnostic parameter in the Boolean grain model.
The method is efficient in discriminating spatial patterns
and selecting structures by the means of conditional subsampling.

%%%%%% Section 7 %%%%%%%%%

\section{Outlook }
\label{7_out}

The calculation of the family of Minkowski--Functionals by using
a Boolean model of penetrable grains and evaluating the local  
(per grain) contributions
to the functionals $W_{\nu\;;\;\nu\ge 1}$ on the surface of the
overlapping union of grains is an effective way of characterizing the local
as well as the global morphology of point processes, as we have just shown.

As a result of the {\it local}
calculation of the contributions to the MF's, this method 
offers a variety of different informations by {\it conditional subsampling} and 
by using various combinations of MF's. 
The latter is necessary as noted earlier \cite{buc95}: the 
Euler characteristic alone is not a preferable tool to discriminate similar
point processes (compare Fig.2 left and right panels in \cite{buc95} and Fig.11).
In addition, geometrical information is needed, which is furnished by 
the other MF's. Only all $4$ measures can (within the class of additive and
motion invariant measures) completely characterize the 
morphology. 

The current numerical code to calculate the MF's performs with reasonable 
storage and CPU time
(we need about 180--250 MB storage and 1--6 h CPU time, depending on 
machine and degree of clustering,
to realize, e.g., a sample of $10^{6}$ points). 

In our statistics group (SFB 375/B3)
we currently work on an alternative way to calculate the MF's: We use a 
grid and calculate the MF's by 
their contributions per grid cell on a sufficiently fine grid spacing. 
The resulting code is comparatively fast and 
storage efficient. Moreover, it implies another advantage: 
in addition to point samples we can consider density-- or temperature fields
(given on a grid) which renders the method applicable to smoothed 
cosmic density fields and microwave background maps.
Smoothing is another key--element of this method which defines one of our 
near future  
perspectives. 

\noindent
We envisage two steps:

\noindent
1. We calculate the MF's of smoothed fields (employing, e.g. Gaussian 
filters). Introducing a second diagnostic parameter (e.g. a density threshold),
this method covers the topology approach by Gott, Melott, Weinberg and 
collab. (see: \cite{mel90} and ref. therein), where we use $4$ functionals
instead of just one.
We here try to restore motion invariance (with respect to the discrete group 
of motions) and additivity,
two important properties to allow for a local characterization of
morphology. This method will also cover the topological/geometrical 
approach on a grid introduced in \cite{mob90}.

\noindent
2. We will combine, so--called {\it Koenderink measures} 
(\cite{kon84},\cite{kon87},\cite{kon91}) 
with MF's.
Koenderink measures provide a set of specific filters which extract different
pattern recogition elements from a smoothed point set. For different 
smoothing lengths the performance of these filters displays an optimum on 
some smoothing scale. 

The combination of these methods will provide a powerful tool for 
the morphological characterization of point processes, density-- and 
temperature fields in scale--space.

A further perspective concerns {\it stereological applications} of MF's
(see, e.g., \cite{wei83}), i.e., the possibility of extracting 3D informations
from lower--dimensional data sets (such as pencil beam surveys).
It will be necessary to analyse smoothed and unsmoothed Minkowski--measures
in the three--dimensional space compared with those in projected distributions.

%\noindent
%With this contribution we release the related software. 
%Please, contact
%$<$ buchert@stat.physik.uni-muenchen.de $>$ to obtain it.

\bigskip\bigskip\noindent
{\bf Acknowledgments:}

\noindent
We would like to thank Rien van de Weygaert for providing his Voronoi
tesselation code, and Simon White for providing his N--body simulation results;
we have benefited from numerous discussions with them and with  
Gerhard B\"orner, Stefan Gottl\"ober, Martin Kerscher, Houjun Mo, Jens 
Schmalzing and Herbert Wagner.

\noindent 
TB is supported by the "Sonderforschungsbereich 375--95 f\"ur
Astro--Teilchenphysik" der Deutschen Forschungsgemeinschaft.

\bigskip
	
%%%%%%%%%%%%%%%%%%%%%%%%%%%%%%%%%%%%%%%%%%%%%%%%%%%%%%%%%%%%%%%%%%%%%%%%
%              Literaturverzeichnis             		       %
%%%%%%%%%%%%%%%%%%%%%%%%%%%%%%%%%%%%%%%%%%%%%%%%%%%%%%%%%%%%%%%%%%%%%%%%

%%%%%%%%%%%%%%%%%%%%%%%%%%%%%%%%%%%%%%%%%%%%%%%%%%%%%%%%%%%%%%%%%%%%%%%%
% Main	End Document			  			       %
%%%%%%%%%%%%%%%%%%%%%%%%%%%%%%%%%%%%%%%%%%%%%%%%%%%%%%%%%%%%%%%%%%%%%%%%

\end{document}